# Predicting the sources of an outbreak with a spectral technique

Vincenzo Fioriti [a,*] and Marta Chinnici [a]

[a]*ENEA Casaccia Laboratories, S. Maria in Galeria, Rome*

Abstract

The epidemic spreading of a disease can be described by a contact network whose nodes are persons or centers of contagion and links heterogeneous relations among them. We provide a procedure to identify multiple sources of an outbreak or their closer neighbors. Our methodology is based on a simple spectral technique requiring only the knowledge of the undirected contact graph. The algorithm is tested on a variety of graphs collected from outbreaks including fluency, H5N1, Tbc, in urban and rural areas. Results show that the spectral technique is able to identify the source nodes if the graph approximates a tree sufficiently.

*Corresponding author
 *Email address*: vincenzo.fioriti@enea.it

*Keywords*: Complex networks, Eigenvalues, Epidemic spreading, Infective diseases

## 1. Introduction

Many infectious diseases spread through direct person-to-person contact. A useful tool in the analysis of epidemic spreading phenomena is the contact network, that represents individuals as vertices and contacts as edges connecting appropriate nodes. More generally, contact networks are understood as a set of heterogeneous relations among nodes and links, where nodes are individuals and links the relations [1, 2]. Today the epidemic spreading paradigm is applied to the malware diffusion [3, 4], the cascading effect in power grids [4], outbreaks of severe infective diseases [3] and rumor spreading in social networks [5, 6, 7]. The formal basis of the analysis is provided by the Graph Theory, that for a number of years has been considered only as a theoretical field of mathematics. Currently, computer calculation power makes it possible [2] to elaborate huge quantity of data and so Graph Theory today is the key to investigate complex networks. The particular problem we are interested in is the identification of the nodes who started the spreading of a disease (or of an information) on a network. This information could be extremely useful to reduce a present or prevent a future outbreak, but from both the mathematical and the practical point of view has been regard as a difficult task [1, 2, 3, 4].

Probably one of the very first attempts to solve directly this problem is due to Shah and Zaman [3]: they provide a framework for the virus spreading on a tree with the SIR (the status of a node is allowed to be susceptible, infected or recovered) framework and construct a maximum likelihood estimator for the source (recently a generalization has been done in [5] for partially unknown infection graphs). To obtain the infection tree from a generic network, they construct a permutation of the infected nodes subject to some ordering constraints due to the forbidden paths and reducing the generic graph to an infection tree. The likelihood of a given source is calculated by adding up the probabilities of all permitted permutations which begin with the source [3]. Unfortunately, the number of permutation is very high and must be reduced. To this end, authors assume the nodes are infected following the fastest spreading, avoiding to consider the longer permutations. They also point out how high degree nodes (hubs) may mislead the algorithm that is prone to select an hub as the origin of the spreading. Thus Shah and Zaman [3] use a statistical tool in order to find, among the several permitted paths of the tree, the most probable unique origin. They claim that simulations performed on small world/scale free and on real networks, such as the U.S. electric power grid, show their estimator finds the target nodes within a few (3 or 4) hops of distance. The computational complexity is $O(N^2)$ for generic graphs, $N$ number of nodes, and $O(N)$ for trees.

A similar work has been done in [6]; authors look for a set (not a single source as in [3]) of sources able to optimize the virus propagation by means of the minimum description length principle using the eigenvectors of the Laplacian matrix of a generic graph. To evaluate the goodness of the solutions they propose two quality measures, claiming their algorithm attains high accuracy in detecting the source nodes and also identifying their number, on some synthetic graphs as well as on the Oregon Internet Autonomous System network. The computational complexity scale linearly with the number of links, although their algorithm is fairly heavy.

In [8] the standard degree, betweenness, closeness, and eigenvector centrality are used to recover the unique starting node of the spreading, because it seems the source node tends to have the highest centrality values. The centrality measurements are applied to Erdos Renyi, scale free graphs and to a real email network. Authors claim their method works very well if the spreading is snowball like and suggest to combine centrality measurements with pattern recognition techniques.

Note that in all the above mentioned papers, including ours, the status of the nodes and the epidemic model are known, that is, the infection graph is known. The general problem of a generic graph with nodes whose status is completely unknown will be object of a future work.

We are interested in developing a deterministic spectral procedure to locate the origins of a disease outbreak, the so-called indexes or primary nodes or their closest neighbors with minimum (actually only the undirected graph topology) information about the spreading. Nodes need not to be individuals: they may be building, cities, geographical areas and links are not restricted to physical interactions. Our procedure is strictly related to the node age, because the node origin of the outbreak should be the oldest of the infection network.

Bearing this in mind, we take advantage of the correlation between eigenvalue and age of node studied by Zhu et al. [9]. They claim that the eigenvalue spectrum of the connectivity matrix, or preferably of the Laplacian matrix, is related closely to the age of nodes of graphs evolving according to non random criteria. The basic requirement is that the network must grow slowly enough: if no evolutionary process has been developed in the past, the method is not applicable and of course the same holds for Erdos Renyi graphs. On the other hand, for graphs following the preferential attachment rule ("rich get richer") the correlation is clear, because the probability for a node to acquire new links is proportional to its degree, therefore a strong correlation between the node degree and its life-time is sure. But real world networks are much more complicated, therefore to find a correlation spectrum-age is not trivial. For a given eigenvalue, the lifetime of the associated eigenvector is the average age of all nodes contained in the vector, weighted by the respective components of the eigenvector. The first step of the Zhu' method [9] is to build the Laplacian matrix [9]:

**L = D – A**.  (1)

where **D** is the degree diagonal matrix and **A** the adjacency matrix ($a_{ij} = 1$ if the link $i$-$j$ exists, 0 otherwise ). The second step is the standardization of each eigenvector components:

$v_i = | v_i / max(v_i) |$, with $i = 1, 2, ... N$.  (2)

The third step is the seniority ranking; nodes with standardized component values larger than a threshold are clustered in a certain age subset and related to the associated eigenvalues, thus the largest eigenvalues is associated to the oldest node and so on. The Zhu's procedure [9] is due to the observation that the eigenvector size in some networks do not seems to increase particularly, while the corresponding (according to the threshold procedure) eigenvalue does. However, no suggestions about the characteristics of evolving networks suitable to be age-analyzed or how to choose the threshold's value are given. Moreover, authors of [3] consider the Laplacian matrix, while we are interested in the adjacency matrix .

In order to ride out these difficulties, we propose to use the *node dynamical importance* introduced by Restrepo et al. [7] to assess the most prominent nodes of a network. It essentially scans the whole graph calculating the reduction of the largest eigenvalue of the adjacency matrix after a node has been removed. We note that a large reduction after the elimination of a node implies the node is relevant to the "aging" of an infection network (that usually, but not always, approximates somehow a tree) and therefore may result to be the origin or topologically close to it. We apply this method to some infection networks from real outbreaks, showing that recovering the origins of an outbreak by the *dynamical importance,* (in this context we will call it the *"dynamical age"*), is possible.

## 2. The dynamical age

We define the *dynamical age* (DA) procedure, as the normalized amount by which $\lambda_m$ (the maximum eigenvalue of the adjacency matrix) decreases if a node *i* is removed and a new $\lambda_m$ is calculated [7]:

$$DA_i = | \lambda_m - \lambda_m^{new} | / \lambda_m . \qquad (3)$$

This parameter describes the dynamical characteristics of nodes. Indeed, suppose you have a synchronized network of oscillator according to:

$$K_c = k_0 / \lambda_m . \qquad (4)$$

where $K_c$ is the coupling synchronization threshold of Kuramoto [10]; thus the maximum eigenvalue is directly related to dynamical properties such as synchronization, percolation and epidemic spreading threshold [7]. Now, note that synchronization can be seen as a strong dependency and that it is also possible to relax it to the so-called generalized synchronization, i.e. a weaker, nevertheless clear, form of dependency [11, 12, 13] among nodes. Hence our choice to use a spectral technique instead of a probabilistic technique, as commonly done in the past, is supported by a wide theoretical and practical background. Most of all, it is a dynamic measurement (as opposed to static parameters such as degree, betweenness, closeness, and eigenvector centrality etc.) able to provide deeper insights of evolving networks [7, 12].

Thus our procedure to identify the node origin of an outbreak, given the infection graph is known, is to calculate $DA_i$ for every node and rank them accordingly, the highest $DA_i$ values indicating the first spreaders of the outbreak. In the next Section we will evaluate the DA algorithm on synthetic graphs to assess its soundness and then on real infection networks.

## 3. The dynamical age and the Barabasi Albert graph

A Barabasi Albert (BA) graph grows from a small number of nodes, adding a node at a time, satisfying the requirements explained in [9]. The probability for a node to be connected to the new node is proportional to the degree according to the preferential attachment model. Each node is sequentially added to the graph and attached to *k* neighbors and nodes that already have many neighbors will have a higher probability of being connected. This means that in a BA model "old" nodes have high degree. Let $d_k$ be the degrees of the *k* highest degree nodes at time *n* in a graph evolving according to the preferential attachment rule, with high probability the largest *k* eigenvalues of the adjacency matrix of this graph will be [18]:

$$\lambda_k(n) = (1 \pm O(1))(d_k)^{1/2}. \tag{5}$$

then the eigenvalues are directly proportional to the degrees and DA will rank them correctly at the first places. The Barabasi Albert like graphs are amenable to be analyzed successfully by the DA algorithm, but if the infection follows a more random pattern, the high degree nodes will have a misleading influence on the DA algorithm, as suggested in [3]. In the real world networks are very different from BA graphs, nevertheless we have the chance to test the *dynamical age* on a well-known mathematical environment just to get an idea about the performance of the DA algorithm. We analyze a 500 and a 1000 nodes BA graph whose oldest nodes are **1, 2, 3, 4**. For the BA500, *dynamical age* classifies (Table 1) node ages as follows: **2, 4**, 11, 15, **1**. Thus the procedure recovers exactly three out of four nodes within an error margin of five results. In order to make a comparison, the performance of a random predictor is calculated: the DA performance of 3 out of 5 correct classifications is reached 486 times out $10^6$ runs by the random guessing (exact ordering is not considered), thus is reasonable to hold the DA classifications are not due to the chance.

**Table 1**

| dyn. age | node | rank |
|---|---|---|
| . | . | . |
| . | . | . |
| . | . | . |
| 0.010216 | **1** | 5 |
| 0.010358 | 15 | 4 |
| 0.010648 | 11 | 3 |
| 0.023519 | **4** | 2 |
| 0.182350 | **2** | 1 |

In Table 2 is shown the *degree* classification of the first ten nodes. Nodes **2** and **4** are the first two larger hubs, but DA recovers also node **1** that is *not* among the most large hubs, see Table 2:

**Table 2**

| degree | node |
|---|---|
| 49 | **2** |
| 30 | **4** |
| 28 | **3** |
| 16 | 16 |
| 12 | 89 |
| 10 | 11 |
| 10 | 13 |
| 9 | 15 |
| 9 | 21 |
| 9 | 27 |

When number of nodes *N* increases, the performance improves; in fact, for *N* = 1000, all the four sources **1**, **2**, **3**, **4** are recovered correctly, as shown in Table 3:

**Table 3**

| dyn. age | node | rank |
|----------|------|------|
| 0.017757 | **3** | 4 |
| 0.021436 | **1** | 3 |
| 0.025797 | **4** | 2 |
| 0.250670 | **2** | 1 |

The DA performance of four out of four correct classifications is reached 66 times out $10^6$ runs by the random guessing. Quantifying the infection age has an important advantage. If two or more of the oldest nodes have very close $DA_i$ values (as node **1** and **4** in Table 3), it means those nodes are coetaneous, therefore it is possible the graph has two or more primary sources.

In [3] authors study a scale free graph with *N* = 400 obtaining an error below four hops on average, thus we conclude that the performance of the DA algorithm on synthetic scale free BA graphs is satisfying.

4. **Results and Discussion**

Some case studies have been taken from the literature to validate our procedure against real data. They are the infection networks of: a pandemic influenza in a U.S. urban area (2006) [15], a pandemic influenza in a school of London (2009) [16], H5N1 outbreak in a rural area in Nigeria (2006) [1], Tbc in Huston (1993-'96) [17] and pandemic influenza in a semirural community in Pennsylvania (2009) [18]. The methodologies used to collect and elaborate data are heterogeneous and errors in their graphs should be expected. An exhaustive data description has been omitted since we are interested in verifying the robustness of the algorithm regardless to data collection methods; in fact, it should be remembered that if the Zhu evolutionary condition and the tree approximation condition are not met, the algorithm's performance will be poor.

**4.1 Pandemic influenza in a U.S. urban area**

The U.S. 2006 pandemic influenza graph (Figure 1a) is composed of 147 nodes that pinpoint a single persons [15]. The exact sequence within four hops form the origin is: **19**; **21**; **24**; **22**, **23**, **61**, **36**; **29**, **37**, **84**, **35**, **79**, **129**, **59**, **75**, **80**, **125**. The DA indicates (oldest first) node **61**, **80**, **24**, **59**, **75**, **79**, **129**, **76**, 100, 60, 95, 86, **36**, **53**, 99, 98, 97, 96, 101, 102, 54, 56, **22**, **21**, **23** (bold nodes are those within four hops). Note that among them only node **24**, **36**, **61** are hubs, therefore DA in this case is not biased by the degree. Within these first 25 results can be found node **21** and **24**, the closest to the actual origin (node **19**, ranked at the 64[th] place). The random guessing gets the same performance 127 times out of $10^6$ runs (ordering is not considered). In Figure 1b the arrows show the sequence of the infections from node to node within four hops from the origin, node **19**.

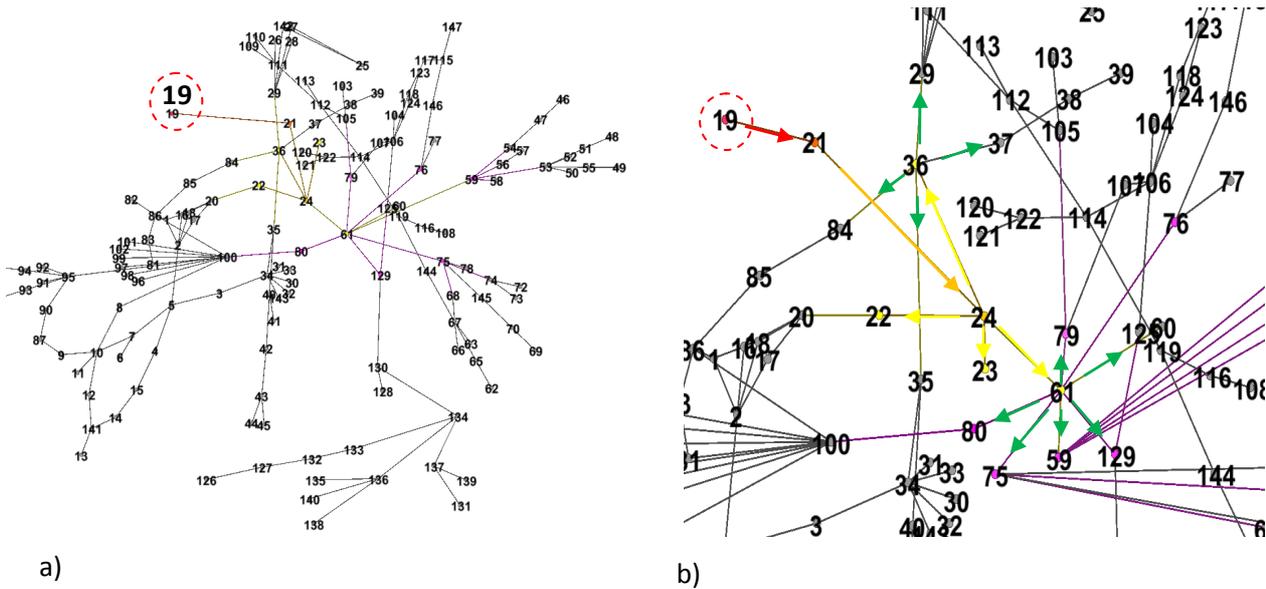

**Fig. 1** a) Graph of the pandemic influenza in a U.S. urban area, $N = 147$, 13 cycles. Node **19** is the origin (red dotted circle) of the outbreak. b) enlargement of the neighbors of the origin; colored arrows indicate the temporal sequence of the infections: first red, then orange, yellow, green (colored nodes are those correctly indicated). There are 16 nodes within 4 hops from the node origin and DA was able to indicate 12 of them amid the first 25 results.

### 4.2 Pandemic influenza in London

This is school-based outbreak of pandemic influenza that occurred in London, United Kingdom, in April 2009. The study included any individual who attended the school as a student or staff member as well as their close contacts; the time of symptom onset was determined, contacts were identified. A case was defined as who showed symptoms and tested positive for virus.

Here DA indicates five nodes **2**, 32, **7, 6**, **1** as the most important spreaders (the exact sequence within two hops is: **1**; **2**, 3 4, 5; **6**, **7**). Actually, node **1** is the source of the pandemic influenza network [16], node **2** is within one hop and nodes **6**, **7** are within two hops from the source. Also in this case the algorithm is not fooled by high degree nodes such as node 32 and 24. The random guessing gets the same performance 2950 times out of $10^6$ runs (ordering is not considered).

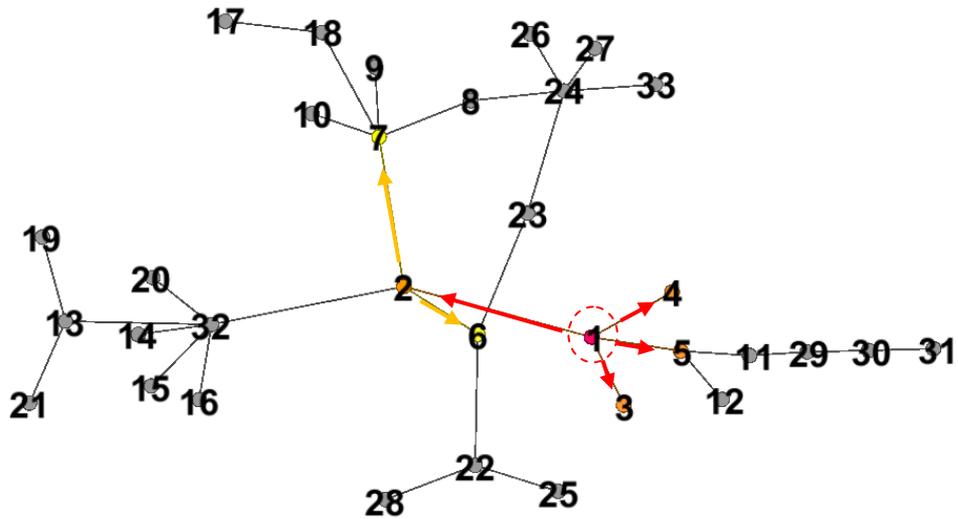

**Fig. 2** Graph of a pandemic influenza in a school of London, $N = 33$, 1 cycle; DA indicates nodes **2,** 32**, 7, 6, 1**; the actual source of the spreading is node **1** (red dotted circle), the next is node **2** (one hop far), both recovered by DA; also nodes **6** and **7** (two hops far from the origin) are identified. Color code of the temporal sequence of infections is: red, orange.

### 4.3 H5N1 outbreak in a rural area in Nigeria

In [1] the model of the outbreak considers contact related and connectivity related networks because mobile people or animals use non mobile connecting networks (roads, water, railroad networks, markets) and such networks are built before they are used by humans and animals. Then the graph of Figure 3 where the links are roads connecting groups of H5N1 cases is not a contact network, rather it is a "connection" network. According to [1], the integration of bio geo temporal data and the connection network generates information usable to optimize control measures and to predict the outbreaks. In Figure 3 the actual sources are nodes **1**, **2**, **3**, **4**, **5**, **6**, **7** (that are considered more or less of the same age). Results from DA are: **3**, 37, 17, 8, 27, 28, **2**, **7**, **1**. The performance is deteriorated with respect to the previous two study cases, because the graph of Figure 3 cannot be approximated to a tree.

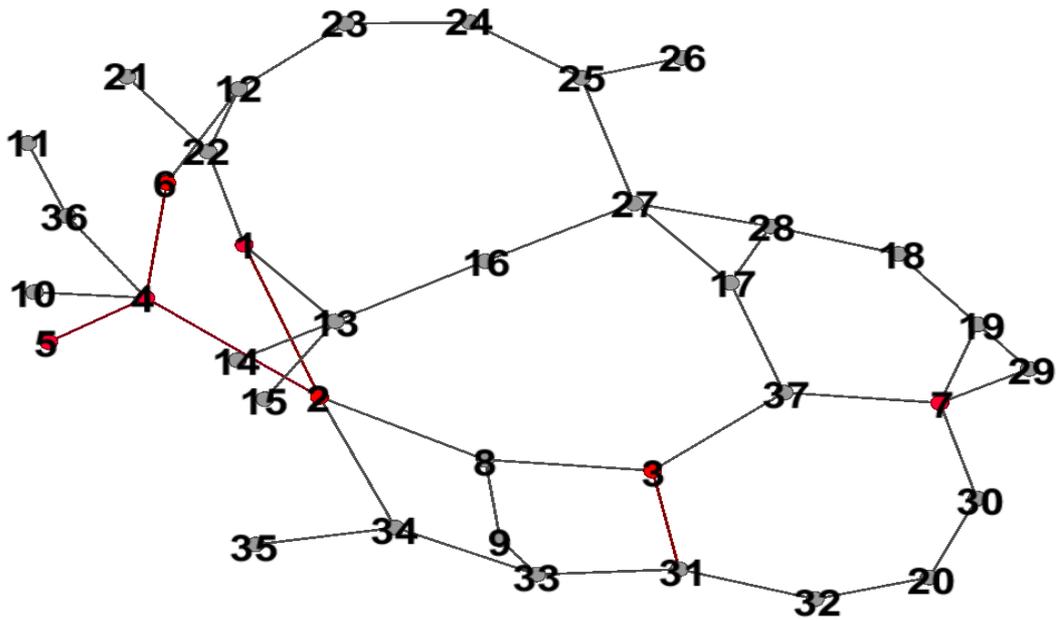

**Fig. 3** H5N1 outbreak in a rural area in Nigeria, $N = 37$, 219 cycles. Here the sources are nodes **1, 2, 3, 4, 5, 6, 7** (red). Results from DA are: **3**, 37, 17, 8, 27, 28, **2, 7, 1**. The random guessing obtains the same performance of DA in the 3.7% of the $10^6$ runs (ordering is not considered).

In Table 4 are showed the first nine $DA_i$ values. Note that node **1, 2, 7** are very close (that is, their ages are similar) and this is consistent with report in [1], thus may be considered as most probable node origins (primary cases).

Table 4

| dyn. age | node | rank |
|---|---|---|
| 0.016169 | **1** | 9 |
| 0.018565 | **7** | 8 |
| 0.020403 | **2** | 7 |
| 0.020531 | 28 | 6 |
| 0.022368 | 27 | 5 |
| 0.022552 | 8 | 4 |
| 0.024560 | 17 | 3 |
| 0.029869 | 37 | 2 |
| 0.030905 | **3** | 1 |

### 4.4 Tbc outbreak in an urban area in Huston

In [17] nodes and links of the outbreak network ($N = 43$) in Figure 4 associate persons, places, objects responsible somehow of the transmission of the infectious agent, therefore the graph do not represents a social network neither approximates a tree. In fact, its performance is poor: 12, 21, 9, 14, 17, 19, 32, **1** are the DA results, while the actual node origin is **1**, immediately after we have node **2, 3, 4, 5** (representing places).

The random guessing in this case obtains the same performance of DA in the 17% of the runs (ordering is not considered), indicating the procedure has lost accuracy.

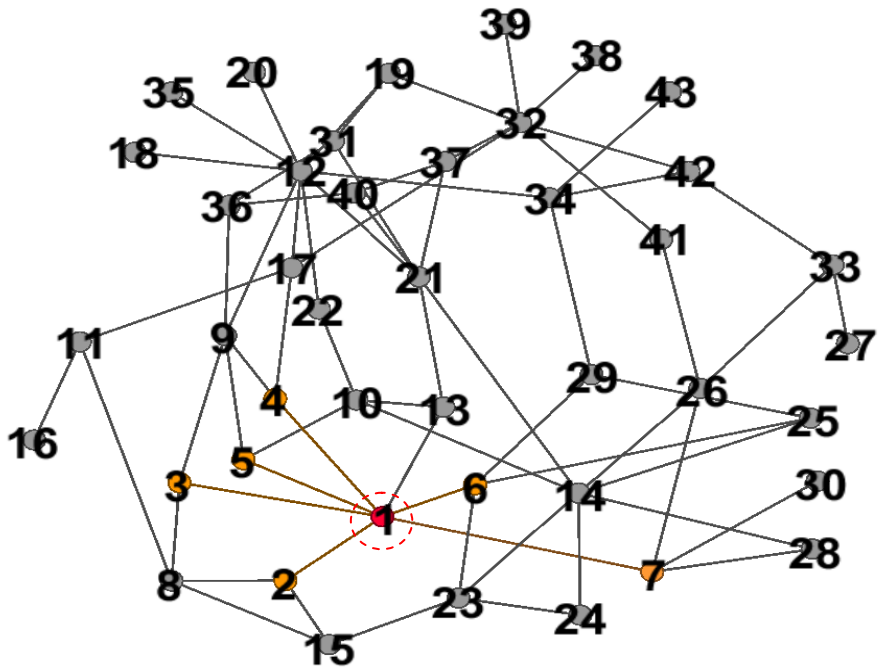

**Fig. 4** Tbc outbreak in an urban area of Huston, $N = 43$, 465952 cycles. The origin is node **1** (red dotted circle), then we have node **2**, **3**, **4**, **5**, **6** (that represents places instead of persons). Results from DA: 12, 21, 9, 14, 17, 19, 32, **1**. The random guessing in this case obtains the same performance of DA in the 17% of the $10^6$ runs (ordering is not considered).

### 4.5 Pandemic influenza in a semirural community in Pennsylvania

In [18] authors reconstruct probabilistically the full infection transmission tree drawing it from the predictive distribution of 2500 trees consistent with the data and, as a consequence, the Pennsylvania flu graph looks like a tree but tends to contain a relevant amount of randomness, departing from the hypotheses of Zhu [9]. Therefore the problem here is different from the Tbc case of Figure 4. Furthermore, data were subject to limitations: many cases were not laboratory confirmed, some were self-reported, some early cases were missed and so on. Therefore the graph ($N = 264$) in Figure 5 is not correct and not timed properly. This last difficulty is particularly relevant, since the whole procedure assumes a rather slow evolution in time of the graph and the *dynamical age* algorithm should not be used if this requirement is not met. As a matter of fact, within the first 84 results node **132** (one hop from the origin) and node **88** (two hops from the origin) are recovered, respectively at the 84$^{th}$ and 66$^{th}$ place. The random guessing obtains the same performance of DA in the 6% of the runs (ordering is not considered). Then the Pennsylvania case is not suitable for the DA analysis and in fact the performance is poor.

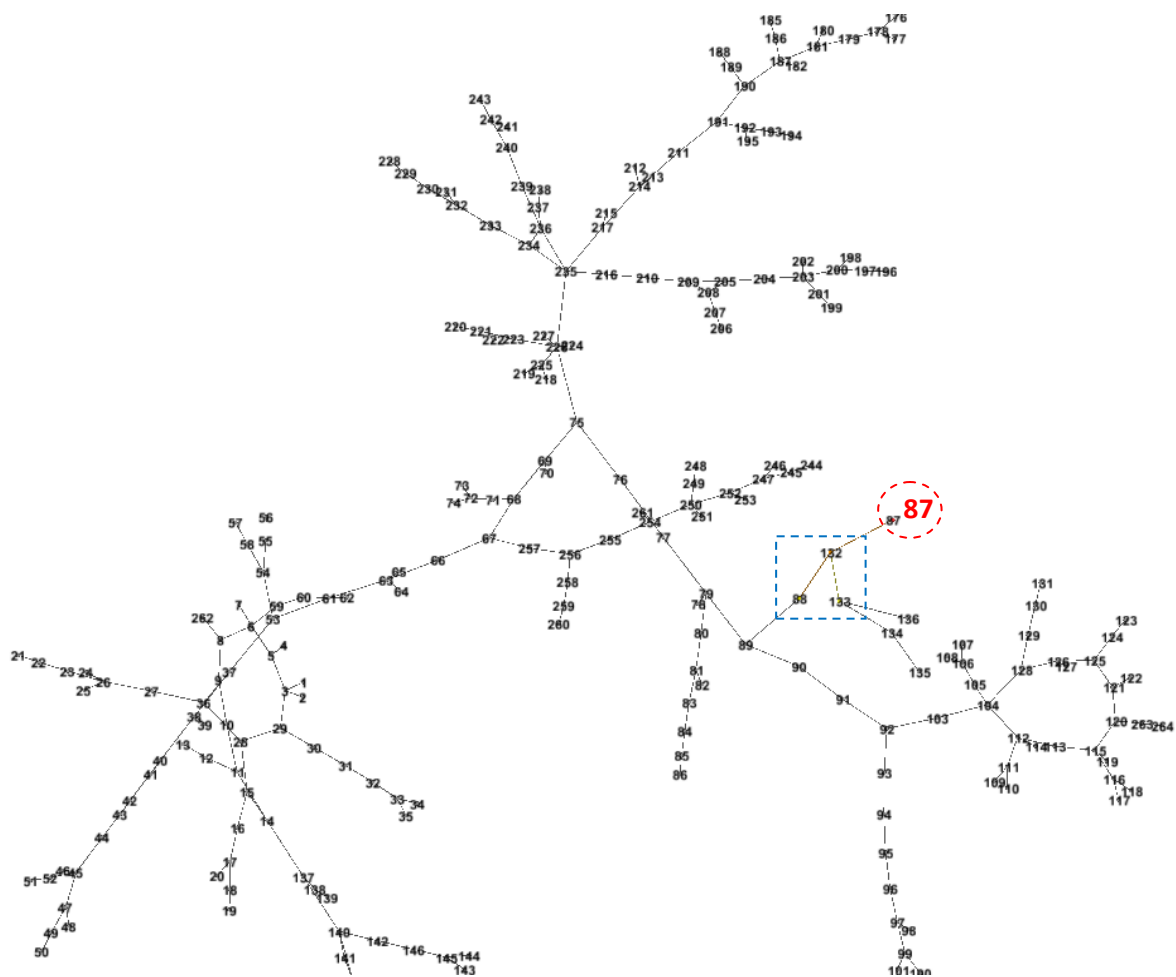

**Fig. 5** Reconstructed tree of a pandemic influenza in a semirural area in Pennsylvania, $N = 264$, 7 cycles. Node **87** is the origin (red dotted circle), followed by node **132** (one hop), **133** and **88** (two hops), shown in the dotted square; they are recovered within the error margin of the first 84 results, hence the performance is low if compared with the previous tests. The random guessing obtains the same performance of DA in the 6% of the runs out of $10^6$ runs (ordering is not considered).

In [3] it was demonstrated that the probability to detect the origin has a threshold effect: for linear trees is zero, while for regular trees it is nonzero. Albeit infection graphs of Figure 1, 2, 3 are not exactly trees, our results confirm their findings since tree like graphs yield a better performance when compared, for example, to the Tbc outbreak (Figure 4). Instead the Pennsylvania outbreak graph of Figure 5 has many linear branches clearly visible and as a consequence, the performance of the DA algorithm is the worst, in accordance with [3] and [9].

In Table 5 are indicated the number of cycles (meaning closed paths) for every graph. The number of cycles suggests roughly how close a graph is to a tree: a graph with many cycles is far from a tree. Best performances in identifying the outbreak origins are obtained in graphs with just a few cycles (excluding the Pennsylvania case that is not suitable to be analyzed by DA, as explained before), confirming the analysis of [3].

The major drawback is clear: the error margin is not predictable in advance, although the correct nodes are indicated within the first 2% - 18%; future work will take account of this problem. Anyway, the *dynamic age* algorithm has proved to be reliable with real outbreaks, when data are usually contaminated by errors and noise, thus the tests should be regarded as very significant.

**Table 5**

| graph | $N$ | cycles | performance |
|---|---|---|---|
| London influenza | 33 | 1 | $1^{th}$ |
| US influenza | 142 | 13 | $2^{th}$ |
| Nigeria H5N1 | 37 | 219 | $3^{th}$ |
| Huston Tbc | 47 | $\sim 10^5$ | $4^{th}$ |
| Penn. influenza | 264 | 7 | - |

## 5. Conclusions

When the graph approximates a tree and for synthetic scale free graphs DA results are very good. The algorithm is remarkably simple and direct allowing the search for multiple sources, although it requires to scan the whole adjacency matrix; however, if this is a serious problem for very large networks, actual contact networks are much more smaller.

It is important to point out how the DA procedure is based on a methodology forming a part of a more general framework devoted to the study of complex networks that has already provided many interesting achievements. Following this research line it is then conceivable to develop a complete spectral model for growing trees describing biological and physical phenomena. Future work will include the extension of spectral techniques to generic graphs when the status of the nodes is unknown.

**Acknowledgements**

V. F. sincerely thanks M. Ruscitti, A. Fioriti, N. Sigismondi, S. Ginevri, C. Iafrate for useful discussions.